\documentclass[twocolumn,showpacs,prb,superscriptaddress]{revtex4}
\usepackage{graphicx}
\usepackage{epstopdf}
\usepackage{longtable}

\begin{document}

\title{Creating and modulating electronic states on noble metal surfaces: ultrathin Ag islands on Si(111)-7$\times$7 as a prototype}

\author{Fang Cheng}
\affiliation{BNLMS, SKLSCUSS, College of Chemistry and Molecular Engineering, Peking University, Beijing 100871, China}

\author{Xiong Zhou}
\affiliation{BNLMS, SKLSCUSS, College of Chemistry and Molecular Engineering, Peking University, Beijing 100871, China}

\author{Yang He}
\affiliation{BNLMS, SKLSCUSS, College of Chemistry and Molecular Engineering, Peking University, Beijing 100871, China}

\author{Qian Shen}
\affiliation{BNLMS, SKLSCUSS, College of Chemistry and Molecular Engineering, Peking University, Beijing 100871, China}

\author{Hailing Liang}
\affiliation{BNLMS, SKLSCUSS, College of Chemistry and Molecular Engineering, Peking University, Beijing 100871, China}

\author{Heng Li}
\affiliation{BNLMS, SKLSCUSS, College of Chemistry and Molecular Engineering, Peking University, Beijing 100871, China}

\author{Jianlong Li}
\email{jlipku@pku.edu.cn}
\affiliation{BNLMS, SKLSCUSS, College of Chemistry and Molecular Engineering, Peking University, Beijing 100871, China}

\author{Wei Ji}
\email{wji@ruc.edu.cn}
\affiliation{Department of Physics, Renmin University of China, Beijing 100872, China}

\author{Kai Wu}
\email{kaiwu@pku.edu.cn}
\affiliation{BNLMS, SKLSCUSS, College of Chemistry and Molecular Engineering, Peking University, Beijing 100871, China}

\begin{abstract}
Various-thickness Ag islands were prepared on Si(111)-7$\times$7 using the one-step deposition at a high substrate temperature. An electronic state centered at -0.40$\sim$-0.15eV versus E$_{Fermi}$, detectable on the surface of the Ag islands thinner than 9 layers, was created by the electronic hybridization between Ag and Si at the Ag-Si interface. Scanning tunneling microscopy/spectroscopy and density functional theory revealed that the thickness of Ag islands determined the strength of the hybridization, leading to a modulation to the energy and intensity of the state on the surface.
\end{abstract}

\received[Dated: ]{\today }
\startpage{1}
\endpage{}

\maketitle

The creation of electronic states on surfaces and the control of them behaving as expected, termed ``manipulation" of surface electronic states, are of great importance in terms of, e.g., surface catalysis\cite{catalysis,catalysis-ma,surface-ji}, nanoelectronics\cite{tautz2006,ji2008}, and the quantum control of electronic states\cite{jvb}.
Although significant progress has been made, most of the solutions relay on dosing adsorbates onto the surface to realize the manipulation of electronic states on that surface in the past practices. However, those adsorbates are most likely a primary obstacle to maximize the effects of such electronic state manipulation, especially to the purpose of surface catalysis, which calls for a method offering the adsorbate-free manipulation of surface electronic states.

Recently, the surface potential of Ag(111) thin films was locally tuned by employing a buried defect at the interface of Ag(111)/Si(111)-7$\times$7\cite{jiangy}. It thus suggests a promising way to realize adsorbate-free manipulation using the interface of hetero-structures, although none electronic state was generated in that system. In this Rapid Communication, we demonstrated creating and modulating an electronic state on the surface of Ag(111) islands deposited on Si(111)-7$\times$7 through the Ag/Si interface. In particular, it was found that a new electronic state sitting at roughly -0.40$\sim$-0.15eV below E$_{Fermi}$ can be routinely probed by scanning tunneling microscopy (STM) and spectroscopy (STS) on the surface of the Ag islands that are as thick as 2.2 nm (about 9 island layers (IL)) on the Si(111) substrate. Whenever an Ag island reaches the thickness of more than 10 IL, the new state becomes invisible eventually and surface standing waves of electrons stands out. Such state, probed on the surface, stems from the hybridization of the underlying Si(111) and the grown Ag(111) islands at the Ag-Si interface, as suggested by density functional theory (DFT) calculations. Both STM/STS and DFT results show that the energy of the state can be shifted and the intensity of the state on the surface is tunable by varying the thickness of Ag islands, due to the electronic hybridization weakening in thicker islands, as elucidated below.

\begin{figure*}[tbp!]
  \includegraphics[bb = 0 0 454 341, scale=0.99]{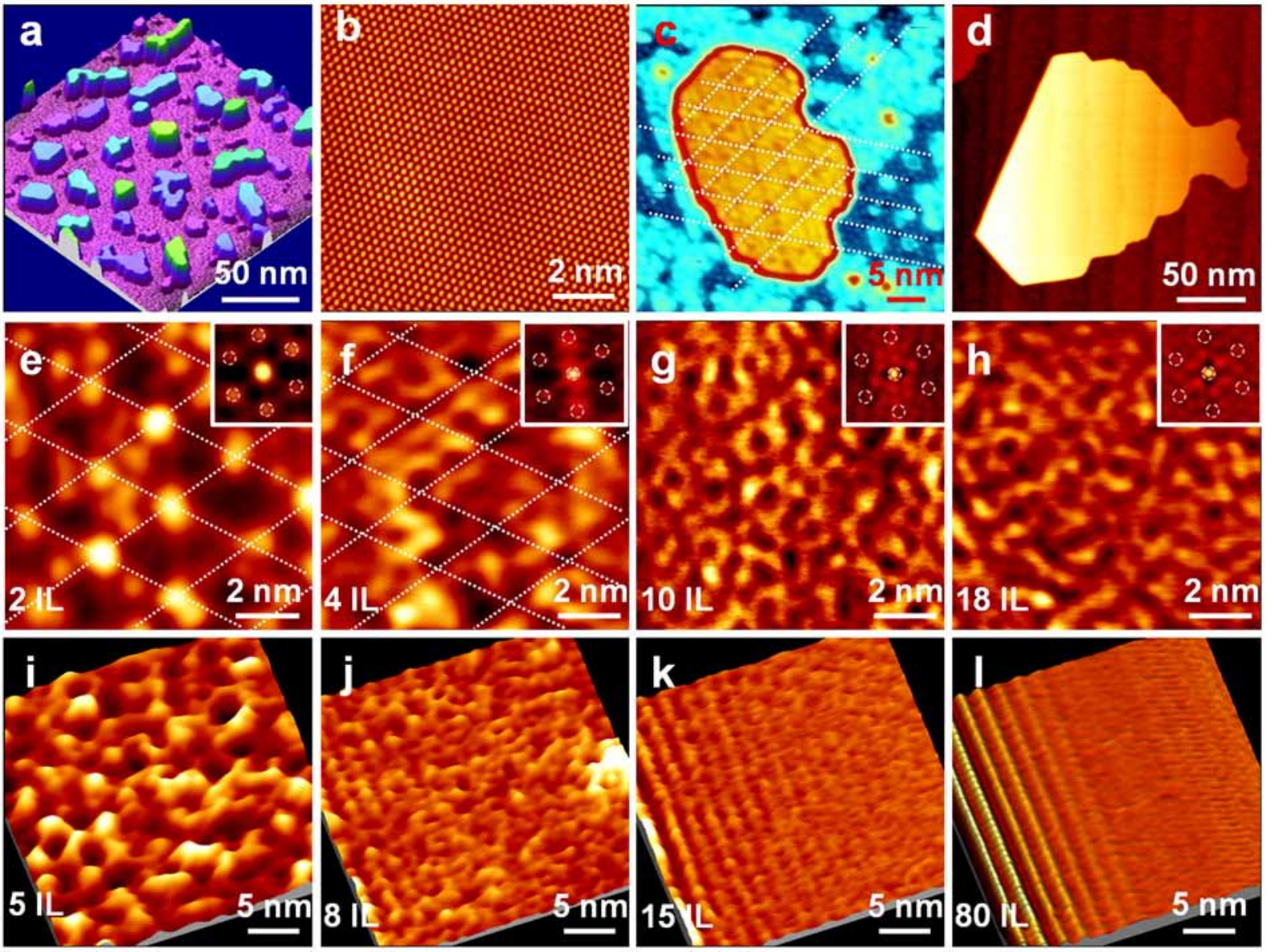}%
  \caption{(Color online). STM images of a typical area showing Ag islands with various thickness (a), an atomic resolution on a 10~IL island (b), a 2~IL-thick island with a partial wetting layer (c), and a wedged island across several Si steps (d) on Si(111)-(7$\times$7). White dotted lines in (c) show the correlation between the Si(111)-(7$\times$7)-like corrugation on island and that on the wetting layer. High resolution STM images of the surface of 2, 4, 10, and 18~IL thick islands were acquired with negative bias voltages (e)-(h). All images were processed by an auto-correlation function ((e)-(h) insets), and the white dotted circles, corresponding to the Si(111)-(7$\times$7) lattice, were marked for better comparison. Four dI/dV mapping images of 5, 8, 15, and 20~IL thick Ag islands were performed at 200~meV (i)-(l).}
  \label{fig:stm}
\end{figure*}

The experiments were carried out in an ultra-high vacuum low temperature STM (USM-1300, Unisoku) with a base pressure below 2.0 $\times$10$^{-10}$ Torr. A clean Si(111)-7$\times$7 substrate was acquired by standard flashing cycles. Afterwards, Ag (Alfa, 99.999\%) was deposited from a K-cell onto the Si(111) substrate held at 120 $^{o}$C. In contrast to the two-step growth mode\cite{two step}, the one-step deposition at a high substrate temperature was chosen to avoid the ``electron growth"\cite{zyzhang} and to obtain a large quantity of Ag islands of various thickness on the same substrate. Then the as-prepared sample was subjected to STM investigations with electrochemically etched tungsten tip at 4.2 K. The tip was cleaned by e-beam bombardment in ultra-high vacuum. The dI/dV spectra and mappings were conducted by adding
a modulation of 10 mV at 1k Hz to the bias voltage via a lock-in amplifier. Herein, the bias voltage in this paper refers to the sample voltage.

Ag islands of various thickness were prepared on Si(111)-7$\times$7 surface, as shown in Fig. \ref{fig:stm}a. A typical high resolution STM image of the surface of those islands was presented in Fig. \ref{fig:stm}b, showing a hexagonal-like periodicity of 2.8 \AA, which indicates that these Ag islands are atomically flat and oriented in (111) direction, consistent with other observations\cite{111}. It was very interesting that a Si(111)-7$\times$7-like corrugation was observable on the top of those islands lower than nine layers, e.g. the two-layer island shown in Fig.\ref{fig:stm}c, although the atomic resolution was harder to achieve than that in higher islands. The lattice seen from the surface corrugation of the island is exactly the same as the lattice suggested from the Si(111)-7$\times$7 surface nearby, as illustrated by white dotted lines, which implies that the 7$\times$7 reconstruction of Si atoms underneath remains unchanged and the corrugation, most likely, represents the information at the Ag-Si interface, i.e. the wetting layer.
Moreover, it is also detectable through Ag islands that the steps at Ag-Si interfaces, see Fig.\ref{fig:stm}d as an example.

STM images at the surface of Ag islands as a function of island thickness were systematically acquired, in order to better understand the observed buried Ag-Si interfacial structure. The observed images, with the thickness up to 8~IL, occasionally 9~IL, share the same periodicity with Si(111)-7$\times$7, while a new periodicity, much smaller than that of Si(111)-7$\times$7, emerges in the images measured from Ag islands higher than 10~IL (inclusive), as shown in Figs. 1e to 1h and insets. These periodicities, however, are only detectable at negative sample biases, and STM images obtained at positive sample bias voltages are rather fuzzy. This substantiates that it's the electronic rather than the geometric effect that dominates in the STM image, and again, the Si(111)-(7$\times$7) lattice is largely preserved at the Ag-Si interface. We thus employed the differential conductance (dI/dV) mapping technique to further investigate the unfilled (positive bias) electronic states
at the surface of Ag islands as a function of island thickness.
In consistent with STM observations, no appreciable periodicity relevant to Si(111)-7$\times$7 was found in any dI/dV mapping image, as shown in Figs 1i to 1l, acquired at a bias of +200 mV. The mapping areas were chosen near island edges so that the surface standing waves could be feasibly detected. These mappings show the gradual appearance of the surface electron standing waves on Ag islands, as the thickness increases from 5~IL to 80~IL. At a thickness of 5~IL, no appreciable surface standing waves could be observed. They, however, faded in when the island was 8 layers in thickness, and became rather clear when the thickness went above 15~IL (Fig. \ref{fig:stm}k and l).

\begin{figure}[tbp!]
  \includegraphics[bb=0 0 253 180, scale=0.8]{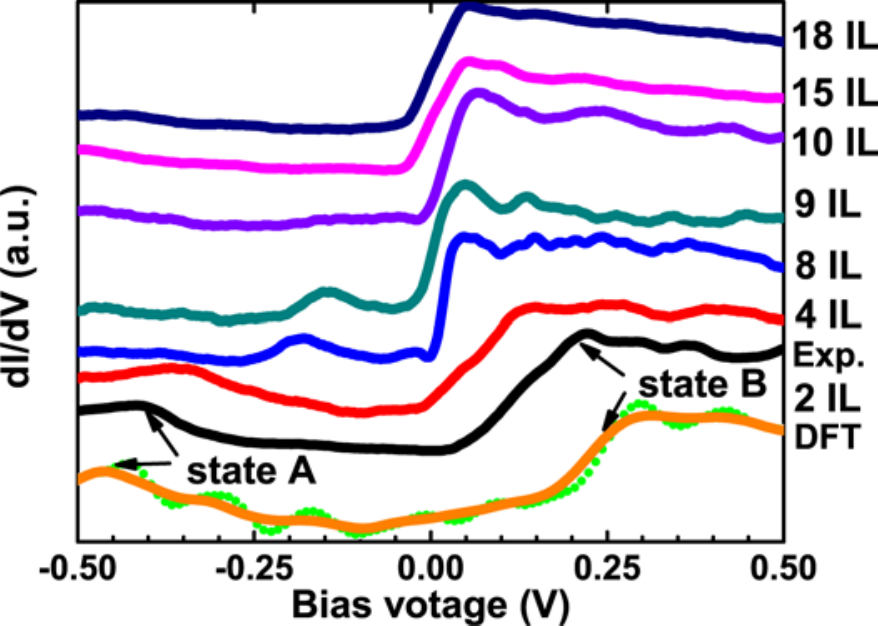}%
  \caption{(Color online). STS spectra for the Ag islands with various thickness on Si(111)-(7$\times$7). The upper six curves are experimental results of islands with their thicknesses labeled at the right side (from 4~IL to 18~IL). The bottom two curves are experimental(Exp.) and theoretical (DFT) spectra of 2-layer islands, as collected together for comparison.}
  \label{fig:sts}
\end{figure}

First-principles DFT calculations were carried out to reveal the underlying physics of the experimental results using general gradient approximation for the exchange-correlation potentials (Perdew-Burke-Ernzerhof)\cite{pbe}, the projector augmented wave method\cite{paw}, and plane wave basis up to 500 eV as implemented in the Vienna {\it ab-initio} Simulation Package VASP\cite{vasp}.
It's fully appreciated that Si and Ag atoms possess radii in a large difference, namely, 5.43\AA~ for Si versus 4.09\AA~ for Ag, which would certainly cause an over 30\% lattice mismatch between Ag(111) and Si(111). It, however, reduces to less than 3\% that the overall strain at the interface in Si(111)-7$\times$7 supercell containing 9$\times$9 Ag atoms in each overlayer, which was experimentally observed. The ``size" of such a supercell with four Si substrate layers and two Ag overlayers, containing roughly 400 atoms, nearly approaches the capacity limit of state-of-art DFT. Nevertheless, a full simulation of it was performed. Lattice constants of Si and Ag were optimized to 5.4739~\AA and 4.1705~\AA. The two bottom Si layers were kept frozen at their bulk positions, other atoms were free to relax until the force on each atom is smaller than 0.005 eV/\AA. Dangling bonds at the bottom of the slab were saturated using H atoms.

The sum of theoretical local density of states (LDOS) of all top-layer Ag atoms (light-green dotted) and its average (orange solid) are shown in Fig.\ref{fig:sts} (2~IL, DFT). It clearly indicates two states, namely, the interface state (state A) at -0.4 eV and the surface state (state B) right above the Fermi Energy (both will be elucidated later). Both states are consistent with the experimental STS spectrum acquired on the top of a 2~IL Ag island (black solid), as shown in the same figure. In terms of state A, its LDOS peak, measured experimentally, moves upward in energy when increasing the island thickness from 2~IL and vanishes after 9~IL, while the step-like feature of state B becomes fairly pronounced from 8 IL and keeps up to 18~IL or even higher.

\begin{figure*}[tbp!]
  \includegraphics[bb=100 10 475 140, scale=0.8]{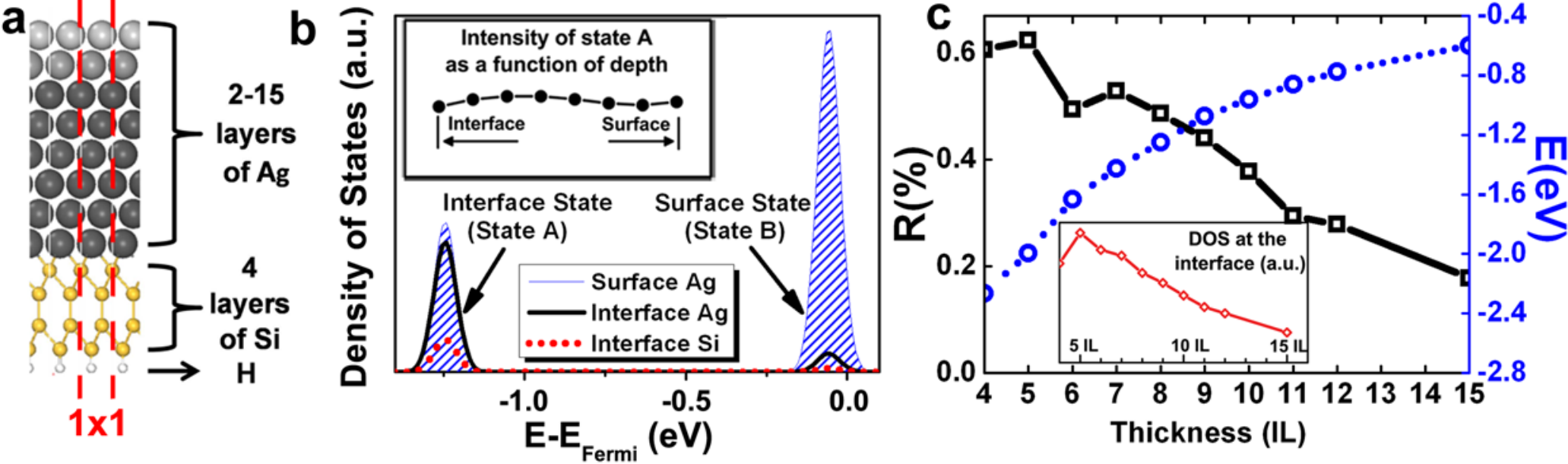}%
  \caption{(Color online). DFT results of Ag islands on Si(111) substrate. (a) Simplified 1$\times$1 model. (b) LDOS of surface (blue solid line with pattern) and interface (black solid line) Ag, and interface Si (red dotted line) atoms on an 8~IL-thick Ag island; and the LDOS of state A of each layer from the surface to interface (inset). (c) Trend of $R$ (black solid line) and $E_{Int}$ (blue dotted line), and the intensity of state A at the interface (inset) as a function of thickness.}
  \label{fig:dft}
\end{figure*}

It is, unfortunately, largely beyond the capacity of state-of-art DFT calculations to do a full simulation modeling Ag islands higher than 2~IL. A simpler model was thus proposed to drastically reduce the computational cost without losing the physics of such a system. Since Ag plays a more crucial role than that of Si in this case, an 1$\times$1 Ag(111) slab containing 4-layer strained (compressed in-plane) Si substrate atoms and 2$\sim$15 layers of Ag atoms was adopted, separated by a vacuum layer of at least 20\AA~, as shown in Fig. \ref{fig:dft}a. The two bottom layers of Si were kept frozen horizontally. All other atoms were free to move until the net force on every atom is less than 0.001 eV/\AA. A k-mesh of 23$\times$23$\times$1 was employed to sample the Brillouin zone. A Gaussian smearing of 0.01 eV was applied in plotting LDOSs, which was integrated from the Gamma point only for clearer showing.

Significant electronic hybridization between Si and Ag at the Si-Ag interface was indicated by the LDOSs of Ag atoms at both surface and interface and a Si atom at the interface of an 8-layer-Ag model, as shown in Fig. \ref{fig:dft}b.
Two states, centered at -1.25 eV and -0.06 eV below the Fermi energy, were found and assigned to the Si-Ag interface and Ag surface states, corresponding to states A and B in Fig. \ref{fig:sts}, respectively.
The emerging of both states is consistent with the full simulation and the experiment. It also suggests that the difference of the interface state positions calculated with Si(111)-(1$\times$1) (Fig. \ref{fig:dft}b) and Si(111)-(7$\times$7) (Fig. \ref{fig:sts}) models stems from the loss of subtle surface atom features, such as the rest-atom, center- and corner-adatoms in the Si(111)-(7$\times$7) model. However, the assessment of the interface and surface states and other physics do validate in the presence of the simplified Si-Ag interface simulated with the Si(111)-(1$\times$1) model.
It's unexpectedly noted that the interface Si atom (red dotted line) not only hybridizes to the interface Ag (black solid line) to form the interface state, but also forms an appreciable wave-function overlap with the surface Ag (blue solid line with patterns) to generate a significant population of the interface state on the surface Ag atoms. Such a population is, most likely, responsible for the observation of the interface corrugation at the surface. Tiny population of the surface state was, however, found at the interface. The layer projected LDOS of each layer from the surface to the interface was shown in Fig. \ref{fig:dft}b inset. The nearly constant intensity of the LDOS as a function of depth clearly indicates the resonance feature of state A, which suggests that the newly formed electronic state is hybridized by an Ag surface resonance and the dangling-bond state at the Si surface.

Figure \ref{fig:dft}c shows the trend of $R$ (black solid line) and the energy of the interface state $E_{Int}$ (blue dotted line) as a function of thickness (T). Variable $R$ is defined as the ratio of the intensities of the interface and surface states at the surface (projected into surface Ag atoms). It clearly indicates that the DOS intensity at the surface from the interface state significantly decreases when increasing the thickness of Ag islands, which agrees with experimental observations that the detection of $7\times7$ periodicity becomes harder in thicker samples.
There was another remarkably result that $E_{Int}$ shifts upwards as the thickness increases, i.e. from -2.0 eV at 4~IL to 0.6 eV at 15~IL. The trend should be valid generalizing it to the real case, i.e. Si(111)-$7\times7$-Ag(111)-$9\times9$, although the absolute energy relative to the E$_{Fermi}$ is much lower in the $1\times1$ model, owing to a larger charge transfer at the interface originated from the neglect of Si(111) surface features in the $1\times1$ model. Indeed, such a trend was experimental observed in STS measurements, as illustrated in Fig. \ref{fig:sts}.

There are two reasonable mechanisms to explain the vanishing of state A at the surface on thicker islands, e.g. due to the band gap (from -0.3 eV)\cite{Ag} of Ag along [111] direction or ascribed to the hybridization weakening. In the former case, if an island goes thicker, the position of state A shifts up approaching the edge of the gap. When it jumps into the gap, the resonating feature of the state will be overwhelmingly suppressed, so that electrons of state A cannot move freely from the interface to the surface, giving rise to the detection of the electrons on the surface unlikely. The fact that state A (centered at -0.15 eV) was found in the gap on a 9~IL island appears to unsupport such mechanism. A 12\% inter-layer expansion was recently reported within the {\it two-step} grown bilayer Ag islands on Si(111)-$7\times7$\cite{prb2010}. Such expansion, ascribed to ``electron growth", can shift $E_{F}$ below the gap. In our case, however, the {\it one-step} method was adopted to avoid ``electron growth".
In terms of the latter case, the intensity of the interface state at the interface was plotted in Fig. \ref{fig:dft}c inset, which points out that the hybridization at the interface, excluding any influence caused between the interface and surface, is weaker when the thickness goes higher. The thicker the island, the higher the energy of the original (before hybridization) first Ag sub-band of the surface resonance state, owing to quantum size effect. From a chemical point of view, such bonding weakening is most likely due to the upward shift of the first sub-band that hybridizes with the Si states to form state A. All these results distinctly support the latter mechanism that the vanishing of the interface state at the surface is due to the hybridization weakening, which is waiting for further experimental confirmation.

In summary, it has been experimentally demonstrated that an interface state, in Ag islands prepared using one-step high-substrate-temperature deposition, can be created by electronic hybridization between an Ag resonance state and dangling-bond states of the Si surface. The wave function of such state, as verified experimentally and theoretically, is capable to pass through the Ag islands, being detected on the topmost surface of the islands. The energy and intensity of the interface state are tunable by varying the thickness of Ag islands, which opens an avenue to realize the adsorbate-free ``manipulation" of surface electronic states. One can reasonably anticipate that molecules or other reactants can be activated by the buried interface to unfold different reaction pathways at the topmost surface, which is very important in catalysis and related processes.

We thank Prof. M.C. Tringides for helpful discussions. This work was jointly supported by NSFC (50821061, 20773001, 20827002, 11004244), MOST (2006CB806102, 2007CB936202, 2009CB929403), BMNSF (2112019), and FRFCU of MOE at RUC (10XNF085). Calculations were done at the PLHPC of RUC and the SSC of Shanghai        .

\end{document}